\documentclass[10pt, twocolumn]{revtex4}
\usepackage{bm,graphicx}
\usepackage[english]{babel}
\usepackage{amssymb, amsthm, amsmath}

\def\be{\begin{eqnarray}}

\def\ee{\end{eqnarray}}

\def\bee{\begin{eqnarray*}}

\def\eee{\end{eqnarray*}}

\newtheorem{thm}{Theorem}

\begin{document}
\title{\bf  Universal quantum computation with little entanglement}
\author{Maarten Van den  Nest}
\affiliation{Max-Planck-Institut f\"ur Quantenoptik, Hans-Kopfermann-Str. 1, D-85748 Garching, Germany.}

\begin{abstract}
We show that universal quantum computation  can be achieved in the standard pure-state circuit model while, at any time, the entanglement entropy of all bipartitions is small---even tending to zero with growing system size. The result is obtained by showing that a quantum computer operating within a small region around the set of unentangled states still has universal computational power, and by using continuity of entanglement entropy. In fact an analogous conclusion applies to every entanglement measure which is continuous in a certain natural sense, which amounts to a large class. Other examples include the geometric measure, $\alpha$-Renyi entropy with $\alpha\geq 1$, localizable entanglement, smooth epsilon-measures, multipartite concurrence, squashed entanglement, and several others. We discuss implications of these results for the believed role of entanglement as a  key necessary resource for quantum speed-ups.
\end{abstract}

\maketitle

\section{Introduction}

Quantum computers are believed to offer exponential computational advantages over classical computers. Understanding  the essential features of quantum physics accounting for this increased power is a fundamental but largely unsolved problem. It is often said that entanglement is the key ingredient that distinguishes quantum from classical computers. Several works have examined the benefits of entanglement, or the absence of such benefits, in quantum computation from various perspectives \cite{Jo97, Go98, Li01, Jo03, Vi03, Or04, Bi04, Da05,  Va06, Jo06, Va07, Ma08, Yo08, Gr09}.
Even though significant progress has been made, the question whether entanglement is a concept that will provide an understanding of quantum computing power in any decisive way (and if so, in which form) is to date an important open question.

The goal of this paper is to examine the role of entanglement as a believed key \emph{necessary resource} for quantum computing.  More particularly we investigate the notion that highly entangled (pure) states should be generated if a quantum computer is to achieve an exponential speed-up (see also \cite{Jo03, Vi03, Jo06, Va06, Va07, Ma08, Yo08}). Perhaps contrary to common intuition, we will show that, throughout any quantum algorithm, the states can remain \emph{weakly} entangled relative to a large class of entanglement measures without significantly compromising the efficiency of the computation. This will hold in particular for the fundamental measure of bipartite pure-state entanglement i.e.  the entanglement entropy. We show that universal quantum computation is possible in the standard circuit model even when, throughout the entire computation, the entanglement entropy of every bipartition is at most $\delta$. Here $\delta$  can be any parameter which scales inverse polynomially with the number of qubits. This means that, as the system size grows, \emph{less} entanglement is required, even tending to zero in the thermodynamic limit.

The proof of the result is elementary. We start by showing that a pure-state quantum computer restricted to operate within a small environment around the unentangled state $|0\rangle^n$ still has universal computational power. Since the entanglement entropy is continuous and equal to zero on product states, its value will be small for any state in such an environment.

As continuity is the key quantity used in the argument, the above result is by no means limited to the entanglement entropy. A fully analogous conclusion applies to every entanglement measure  which is continuous in a certain natural sense. This includes many commonly considered (bipartite and multipartite) measures such as the geometric measure \cite{Sh95}, $\alpha$-Renyi entropies with $\alpha\geq 1$, localizable entanglement \cite{Ve04}, relative entropy of entanglement \cite{Ve97}, squashed entanglement \cite{Ch04}, multipartite concurrence \cite{Ca04},  smooth $\epsilon$-measures \cite{Pi08}, $n$-tangle \cite{Wo01}, and others. In short, having only small amounts of these types of entanglement does not provide any obstacle for universal quantum computation.

\section{Entanglement as a resource for quantum computation}\label{sect_background}

Before stating our main results, we discuss some background related to the assertion that entanglement is a resource required for quantum computation.

Certainly, some nonzero pure-state entanglement is necessary for quantum computation in the sense that quantum circuits which remain in a pure unentangled state throughout the computation cannot yield exponential quantum speed-ups \cite{Jo03}. This basic fact however provides little insight in the potential role of entanglement in quantum algorithms. More useful is the intuition that, as the system size increases, growing amounts of entanglement should be generated in order for a quantum algorithm to achieve an exponential quantum speed-up. Some works have provided precise quantitative statements of this kind \cite{Jo03, Vi03,  Jo06, Va07, Ma08, Yo08}). For example, unless quantum and classical computation have equal power,  large amounts of \emph{Schmidt-rank} entanglement must be generated in a universal quantum computer \cite{Vi03}. More precisely, for an $n$-qubit system, let $\chi_{A, B}$ denote the logarithm of the Schmidt rank of the bipartition $(A, B)$ of the $n$ qubits (see also appendix \ref{sect_app_entanglement}).

\begin{thm}[\cite{Vi03}]
Consider an $n$-qubit quantum circuit acting on a standard basis input and followed by a standard basis measurement. If in every step of the computation  $\chi_{A, B}=O(\log n)$ for all bipartitions $(A,B)$, then this circuit can be simulated  classically in poly$(n)$ time.
\end{thm}
Such a result  might suggest a general conclusion along the lines of ``If the entanglement is not high enough, then a quantum speed-up cannot occur.'' Recall however that  \emph{the} entanglement of a multipartite state does not exist. There are infinitely many  notions of entanglement, all of which are valid measures in their own right but many of which are inequivalent. In particular there exist scenarios where the same quantum state is highly entangled relative to one measure whereas it is only slightly entangled relative to another one (and indeed our results will provide a clear illustration of this).

Given the richness of the concept of entanglement, it is intriguing that to date only few concrete entanglement measures have been shown to be required for quantum speed-ups in a sense analogous to theorem 1. Furthermore, the measures considered so far have very similar definitions, based on the Schmidt rank or closely related concepts \cite{Jo03, Vi03,  Jo06, Va07, Ma08, Yo08}.  In addition, it is known that these results are by no means complete i.e. they cannot account for all families of efficiently simulatable quantum circuits; the Gottesman-Knill theorem \cite{Go98} is a well-known counterexample \footnote{It is very likely that other classical simulation results such as e.g. \cite{Va02}  are not covered by theorem 1 or other low-entanglement simulation schemes either, but this has not been explicitly verified.}.

In other words, even though entanglement is such a vast paradigm, our intuition about its role as a necessary resource for computation seems to be drawn from a handful of analgously defined measures. It is therefore natural to ask: which other types of entanglement must be generated in large amounts to allow for nontrivial quantum computing power?  What are the features of this class,  do generic measures belong to it? etc. This is the main problem addressed in this paper. Anticipating the results, we will in fact show that a number of widely used measures do \emph{not} belong to this class.

Finally, there are important distinctions between the roles of pure- and mixed-state entanglement in computation.  Since unentangled mixed states are nontrivial objects, even the very basic question whether computations involving exclusively fully separable mixed states can be efficiently simulated classically is to date unanswered.  It is well possible that such ``separable computations'' may yield quantum computational benefits \cite{Jo03} and several works exist providing evidence supporting this belief \cite{Li01, Bi04, Da05}. The present work is somewhat similar in spirit as we show that, even for pure states, small amounts of entanglement (relative to various measures) suffice for universal quantum computation.

\section{Entanglement entropy}\label{sect_entanglement_entropy}

Next we show that universal quantum computation is possible even when the entanglement entropy remains small in every step of the computation.

We consider the standard pure-state circuit model. The input is the $n$-qubit state $|0\rangle^n$ and circuits consist of poly$(n)$ elementary unitary gates acting on at most $d$ qubits for some constant $d$. In this work we will in fact always have $d\leq 3$. The computation is followed by a standard basis measurement on the first qubit. The complexity class BQP (bounded-error quantum polynomial time) represents all decision problems that are efficiently solvable on a quantum computer with bounded error probability. In this paper we use the term ``universal'' in a computational sense: a computational model is called universal for quantum computation if it has the power to solve every problem in BQP in polynomial time.

For any $\epsilon>0$ the set ${\cal S}_{\epsilon}$ consists of all $n$-qubit states $|\psi\rangle$ which are $\epsilon$-close to $|0\rangle^n$ in trace distance. A family of parameters $\{\epsilon_n\}$, where $\epsilon_n>0$ and $n=1, 2, \dots$,  is said to be polynomially small if $1/\epsilon_n = O(p(n))$ for some polynomial $p(n)$. We denote by QC$_{\epsilon_n}$ a restricted quantum computer where only those quantum circuits are allowed for which the state of the $n$-qubit register belongs to ${\cal S}_{\epsilon_n}$ in each step of the computation, for every $n$. Henceforth we will simply denote $\epsilon_n\equiv \epsilon$.

\

\noindent {\bf Observation 1} {\it Consider any  polynomially small $\epsilon$. Then QC$_{\epsilon}$ has universal computational power. That is, having access to such a computer allows one to solve every problem in BQP in polynomial time. }

\

This claim is proved as follows. Let ${\cal C}$ denote an arbitrary polynomial-size $m$-qubit quantum circuit composed from a universal gate set (say CNOT gates, Hadamard gates and $\frac{\pi}{8}$-phase gates).  Suppose that ${\cal C}$ acts on the input $|0\rangle^m$ and is followed by measurement of the first qubit in the computational basis. Let $p$ denote the probability of measuring 1.  Then the following problem is well-known to be BQP-complete: given the  promise that either $p\geq 2/3$ or $p\leq 1/3$, determine which of these possibilities is the case. Next we show that this BQP-complete problem can be solved efficiently by means of a transformed quantum circuit operating within the QC$_{\epsilon}$ model. The input of the new circuit is the $n$-qubit state $|0\rangle^n$ where $n:=m+1$. First a single-qubit rotation on qubit $m+1$ is performed to generate the state \be\label{product_state} |0\rangle^m\otimes(\sqrt{1-\epsilon}|0\rangle + \sqrt{\epsilon}|1\rangle).\ee Then each gate in the circuit ${\cal C}$ is applied controlled on qubit $m+1$ being in the state $|1\rangle$. Hence the resulting circuit consists of gates acting on at most three qubits. Letting ${\cal C}_t$ denote the product of the first $t$ gates in ${\cal C}$, it follows that after $t$ gates, the quantum register is in the state \be \label{step_t}|\psi_t\rangle = \sqrt{1-\epsilon}\ |0\rangle^{m}\otimes |0\rangle + \sqrt{\epsilon}\ {\cal C}_t|0\rangle^m\otimes |1\rangle.\ee After all gates have been applied, a standard basis measurement on the first qubit is performed. The probability $q$ of measuring 1 is given by  $q = \epsilon p$. Repeating the computation poly$(n)$ times allows to estimate $q$ with accuracy $1/$poly$(n)$. Since $\epsilon$ is polynomially small, this allows one to determine in polynomial time whether $p\geq 2/3$ or $p\leq 1/3$.
Finally, remark that the overlap between $|\psi_t\rangle $ and $|0\rangle^{n}$ is $\sqrt{1-\epsilon}$ for every $t$. Therefore the entire computation operates within ${\cal S}_{\bar\epsilon}$ with $\bar\epsilon= \sqrt{\epsilon}$ (see appendix \ref{sect_app_basic}). The result now readily follows.

\

\noindent {\bf Observation 2} {\it Consider an arbitrary polynomially small $\delta\equiv \delta_n$. Then it is possible to efficiently solve every problem in BQP even when, throughout the entire computation, the entanglement entropy of every $n$-qubit state is at most $O(\delta_n)$ for every bipartition.}

\

To prove observation 2, consider an $n$-qubit state $|\psi\rangle$ in ${\cal S}_{\epsilon}$ where $\epsilon$ will be determined later. Consider an arbitrary bipartite split $(A, B)$ of the system. Let $\rho^A$ and $|0\rangle^A$ denote the states obtained from $|\psi\rangle$ and $|0\rangle^n$, respectively, by tracing out all qubits in $B$. The entanglement entropy $E^{A, B}(|\psi\rangle)$ is given by the von Neumann entropy $S(\rho^A) = - \mbox{ Tr} \rho^A \log \rho^A$. We now recall  the following continuity property \cite{Fa73}. Let $\rho$ and $\sigma$ be two arbitrary density operators on a $d$-dimensional Hilbert space and denote by $T$ their trace distance.  Then, as long as $T\leq 1/(2e)$, one has \be\label{fannes} |S(\rho)-S(\sigma)|\leq 2T \log_2(d) - 2T \log_2 (2T ).\ee  Since  $T(|\psi\rangle, |0\rangle^n) \leq \epsilon$ and since the trace distance is contractive, this implies that  $T(\rho^A, |0\rangle^A) \leq \epsilon$. Using (\ref{fannes}) and the fact that $|0\rangle^A$ has zero entropy, it follows that \be E^{A, B}(|\psi\rangle)= S(\rho^A)\leq 2\epsilon |A| - 2\epsilon \log_2 (2\epsilon),\ee for every $\epsilon\leq 1/(2e)$, where $|A|$ denotes the number of qubits in $A$. It follows that, given any  polynomially small $\delta$, there exists a suitable polynomially small $\epsilon$   such that $E^{A, B}(|\psi\rangle)=  O(\delta)$. Combining this last property with observation 1 proves the result.

\

Observation 2 is in sharp contrast with theorem 1. In particular, whereas quantum circuits generating logarithmic amounts of schmidt-rank entanglement can be simulated efficiently classically, $\delta$-amounts of entanglement entropy suffice for universal quantum computation---note that $\delta$ \emph{decreases} with growing $n$.

It was previously known that there exist quantum circuits (e.g. Clifford circuits) generating states with large entanglement entropies but which can nevertheless be simulated efficiently classically \cite{Go98}. Conversely,  small entanglement entropies do not form an obstacle for universal quantum computation owing to observation 2. In short,  large amounts of entanglement entropy are neither necessary nor sufficient for quantum-speed-ups.

In the proof of observation 2, the decrease in entanglement entropy comes with an increase in the number of runs of the computation, the latter scaling as a polynomial in $1/\epsilon$ for computations operating within ${\cal S}_{\epsilon}$.  Remark that the ``integrated'' entanglement, i.e. the value obtained by summing  $E^{A, B}$ over all runs, may be large. Nevertheless at no time in the computation is a state prepared with entanglement entropy larger than $O(\delta)$.

\section{Continuous measures}\label{sect_continuous}

The only properties of the entanglement entropy $E^{A, B}$ used to prove observation 2 are (a) this function vanishes on the product state $|0\rangle^n$ and (b) it is sufficiently continuous in the following sense: for every polynomially small $\delta$ there exists a polynomially small $\epsilon$ such that $E^{A, B}(|\psi\rangle)= O(\delta)$ for all $|\psi\rangle\in {\cal S}_{\epsilon}$.
This type of continuity is rather natural and thus exhibited by various other well known entanglement measures; this includes bipartite measures such as $E^{A, B}$ as well as  various true multipartite measures. As a result, observation 2 can readily be generalized:

\

\noindent {\bf Observation 3} {\it Consider an arbitrary polynomially small $\delta$. Then it is possible to efficiently solve every problem in BQP even when, throughout the entire computation, each of the following measures is  $O(\delta)$.

\vspace{2mm}

(a) $\alpha$-Renyi entropy for every bipartition, for all $\alpha\geq 1$;

(b) Geometric measure \cite{Sh95};

(c) Relative entropy of entanglement \cite{Ve97};

(d) Squashed entanglement \cite{Ch04};

(e) Localizable entanglement of every qubit pair \cite{Ve04};

(f) Multipartite concurrence \cite{Ca04};

(g) $n$-Tangle \cite{Wo01}.

}

\

\noindent Definitions of these measures and a proof of observation 3 are given in the appendices. The crux of the argument is that all quantities are continuous in the sense described above.

The list of entanglement measures in observation 3 can be made considerably longer. Whereas we will not attempt to make this list complete, it is interesting to note that analogous conclusions to observation 3 can be reached in one go for generally defined  \emph{families} of measures. We give a sketch of the results, details are given in the appendices.

A first natural example is the family of distance measures, which have the form \be D(|\psi\rangle) = \inf \ \{d(|\psi\rangle, \sigma): \sigma \mbox{ is unentangled}\}.\ee  Here $d(\cdot, \cdot)$ is some  notion of distance and the minimization is either taken over all pure or mixed separable states $\sigma$, depending on the definition of $D$. Thus $D$ measures the distance to the nearest unentangled state. Examples are the geometric measure and the relative entropy of entanglement. Clearly, any other distance measure can be added to observation 3 as long as the measure $d$ is sufficiently well-behaved; that is,  by choosing $\epsilon$ polynomially small one can ensure that the $d$-distance between any $|\psi\rangle$ in ${\cal S}_{\epsilon}$ and $|0\rangle^n$ is at most $\delta$.

A second example is the family of epsilon-measures \cite{Pi08}. If $E$ is an entanglement measure and $\epsilon>0$ then the associated $\epsilon$-measure is defined as \be E_{\epsilon}(|\psi\rangle) = \inf\ \{E(\sigma): \sigma \mbox{ s.t. } T(|\psi\rangle, \sigma)\leq \epsilon\}\ee where $\sigma$ may generally be a mixed state and where $T(\cdot, \cdot)$ denotes the trace distance. Thus $E_{\epsilon}$  measures the minimal entanglement guaranteed to present in an $\epsilon$-ball around $|\psi\rangle$. This construction gives a technique to obtain a smooth function $E_{\epsilon}$ even when the original measure $E$ is not \cite{Pi08}. Observation 1 immediately implies that universal quantum computation can be achieved with \emph{zero} $\epsilon$-entanglement, for \emph{every} underlying entanglement measure $E$ and for every polynomially small $\epsilon$. An interesting example is the Schmidt rank: universal quantum computation is possible with zero Schmidt rank $\epsilon$-measure for all bipartitions---note the sharp contrast with theorem 1.

Beyond $\epsilon$-measures, a similar conclusion holds for other ``smoothed'' entanglement measures which include an accuracy parameter in their definition. For example the one-shot bipartite entanglement cost under LOCC with accuracy $\epsilon$ \cite{Bu11} will similarly be zero.

A third family regards measures related to polynomial functions in the entries of a state. Consider quantities of the form \be\label{polynomial} \langle \psi|^{\otimes k} A |\psi\rangle^{\otimes k}\quad\mbox{ and }\quad \langle \psi|^{\otimes k} A |\psi^*\rangle^{\otimes k}\ee where $k = $ poly$(n)$, where $A$ is an $nk$-qubit operator and where $|\psi^*\rangle$ denotes the complex conjugate of $|\psi\rangle$ in the standard basis. The quantities (\ref{polynomial}) define  polynomials in the coefficients of $|\psi\rangle$ and their complex conjugates.   Several entanglement measures are given by expressions of the form (\ref{polynomial}) or as simple functions thereof. Consider e.g. the multipartite concurrence and the $n$-tangle, as well as the general family of comb-based measures \cite{Os05}.  Another class related to (\ref{polynomial}) with $k=1$ regards witness-based measures of the form \be\label{E_witness} E_{\cal C}(|\psi\rangle) = \max \{ 0, -\max_{W\in \ {\cal C}} \ \langle\psi|W|\psi\rangle\},\ee where ${\cal C}$ denotes a subfamily of entanglement witness operators \cite{Br05}.

Provided that the operator norm of $A$ scales at most polynomially with the number of qubits, every quantity of the form (\ref{polynomial}) is sufficiently continuous for our purposes. As a result, for a large class of entanglement measures based on such quantities will a result similar to observation 3 hold. This is e.g. the case for the multipartite concurrence, $n$-tangle and witness-based measures $E_{\cal C}$ for which the operator norm of each $W\in {\cal C}$ scales at most polynomially with the system size.

\section{Insufficiently-continuous measures}\label{sect_discontinuous}

Although the conclusions of sections \ref{sect_entanglement_entropy} and \ref{sect_continuous} apply to a wide spectrum of measures, a few of them are not covered and it is instructive to consider examples.

A first example regards \emph{discrete} measures such as the Schmidt rank. Obviously, the continuity argument used to prove observation 2 does not carry over. Note that this is in nice agreement with theorem 1. In this context it is interesting to remark the following. Besides theorem 1, there exist (a few) other classical simulation results of the general form \emph{``If entanglement of the type $E$ throughout a computation is small, then an efficient classical simulation exists''} for certain measures $E$, see e.g. \cite{Jo03, Va07}.  Interestingly, these results are also formulated using discrete measures $E$. The results of section \ref{sect_continuous} shed light onto why this is the case; in particular they clarify why an analogous classical simulation result has not been found for continuous measures such as e.g. the entanglement entropy or the geometric measure.

A second interesting class of examples  regards certain $\alpha$-Renyi entropies. For any $\alpha>0$ with $\alpha\neq 1$ define \be S_{\alpha}(\rho)= \frac{1}{1-\alpha} \log \mbox{Tr}( \rho^{\alpha})\ee where $\rho$ is a density operator on a $n$-qubit system. The Renyi entropies generalize the von Neumann entropy $S(\rho)$ in the sense that $S=\lim S_\alpha$ when $\alpha\to 1$. Remark that the maximal value of the Renyi entropy is $n$ independent of $\alpha$, so that the scale of what is considered ``small'' $\alpha$-entropy is the same for all $\alpha$. The pure-state $\alpha$-entanglement   entropy $E_{\alpha}^{A, B}$ is obtained in the natural way, i.e. by considering a bipartition $(A, B)$ and by defining $E_{\alpha}^{A, B}$   to be the $\alpha$-Renyi entropy of the reduced density operator of subsystem $A$.

All measures $E_{\alpha}^{A, B}$ with $\alpha\geq 1$ are contained in observation 3. In particular, for every polynomially small $\delta$ there exists a polynomially small $\epsilon$ such that $E_{\alpha}^{A, B}\leq \delta$ for every $|\psi\rangle\in {\cal S}_{\epsilon}$. In contrast,  such result does not hold for any $\alpha<1$. The following example shows that for every such $\alpha$ there exists $|\psi\rangle$ which is \emph{exponentially close} to $|0\rangle^n$ but where, nevertheless, $E_{\alpha}^{A, B}$ scales linearly with $n$ for some bipartitions.

To this end, denote $\beta:= [1-\alpha]/2\alpha$, consider an $n$-qubit system with even $n=2m$  and set $\epsilon:= 2^{-\beta m}$. Now consider the state \be|\psi\rangle = \sqrt{1-\epsilon}|0, 0\rangle + \sum \sqrt{\frac{\epsilon}{2^m-1}} |x, x\rangle\ee where the sum is over all $m$-bit strings $x\neq 0$ and where $|x, x\rangle$ denote $2m$-qubit computational basis states in the usual sense. The trace distance between $|\psi\rangle$ and $|0, 0\rangle$ is $\sqrt{\epsilon}$ and thus exponentially small in $n$. On the other hand, a direct calculation shows that the $\alpha$-entanglement entropy w.r.t. the natural bipartition of the system into two subsets of $m$ qubits is scaling as $m/2$ i.e. \emph{linearly} with the total number of qubits.

Thus we cannot include any  $E_{\alpha}^{A, B}$ measure with $\alpha< 1$ in observation 3 since the continuity argument does not apply. We believe that this is not a shortcoming of the argument but rather a genuine difference in behavior for the different regimes of  $\alpha$. In fact a result analogous to theorem 1 might well hold for $\alpha$-entropies with $\alpha<1$. Evidence for this is given in Ref. \cite{Ve04} where it was shown that every $n$-qubit state  where $E_{\alpha}^{A, B}=O(\log n)$ for certain bipartitions can be approximated to accuracy $\epsilon$ by a matrix product state of bond dimension $D=$ poly$(n, \frac{1}{\epsilon})$.

\section{Mixed states}\label{sect_mixed}

Whereas the main focus of this work regards pure-state computations, here we briefly consider mixed states.
First, an analogue of observation 1 rather trivially holds if the state of the quantum register is allowed to be mixed. Letting $\epsilon$ be any polynomially small parameter, it is easy to show (see also \cite{Li01}) that every quantum circuit can be efficiently simulated by  a computation where in each step the state has the form \be \epsilon |\psi\rangle\langle \psi| + (1-\epsilon)M \ee for every polynomially small $\epsilon$, where $M=I/2^n$ denotes the fully mixed state on $n$ qubits. Every state of this type (sometimes called \emph{pseudo-pure} state) is easily seen to be $\epsilon$-close to $M$ in trace norm, thus immediately yielding an analogue of observation 1. The ensuing argument naturally carries over as well: every entanglement measure $E$ which is sufficiently continuous will take on small values on every pseudo-pure state, so that universal quantum computation can be achieved with small amounts of entanglement of the type $E$.

In addition, regardless of continuity properties, every \emph{convex} measure $\bar E$ will also be at most $\epsilon \bar E(|\psi\rangle)$ on any pseudo-pure state. This value can be made polynomially small by a suitable choice of $\epsilon $ as long as $\bar E(|\psi\rangle)\leq $ poly$(n)$ for all $n$-qubit states $|\psi\rangle$. Thus also for such measures universal quantum computation can be achieved with small amounts of entanglement. An interesting example is the mixed-state measure $\chi^{A, B}_m $ associated with   $\chi^{A, B}$ via convex roof: we find that universal mixed-state quantum computation is possible even when, throughout the computation, $\chi^{A, B}_m \leq \delta$ for every bipartition, for any polynomially small $\delta$. It is again interesting to note the sharp contrast between this property and theorem 1. In particular, the fact that the Schmidt rank is discrete, which prevented us from adding it to observation 3, does not provide an obstacle in the present context.  In fact, a similar argument can be applied to the convex roof measures defined by other discrete measures, such as the $p$-blockedness \cite{Jo03} and Schmidt measure \cite{Ei01}, as well as $\alpha$-entropies for \emph{all} $\alpha$.

In summary, using the concept of pseudo-pure states one readily finds that universal quantum computation is possible with limited amounts of mixed-state entanglement of certain type, essentially determined by continuity or convexity of the entanglement measure. This conclusion is however less surprising than observations 2 and 3,
since the latter regard pure-state computations.

\section{Discussion}\label{sect_discussion}

We have shown that every quantum circuit can be simulated by a circuit operating entirely within a (pure-state) $\epsilon$-neighborhood ${\cal S}_{\epsilon}$ of the trivial state $|0\rangle^n$. This comes at the cost of an increased number of repetitions of the computation; as long as $\epsilon$ is polynomially small, the required number of runs is polynomial and the overall computation is efficient. This basic mechanism allows one to decrease several forms of entanglement generated during a quantum circuit while maintaining computational universality. In particular, every entanglement measure which is sufficiently continuous will take on small values on ${\cal S}_{\epsilon}$ and will hence remain small throughout the computation---its value even tending to zero as the system size increases. A number of commonly used entanglement measures are continuous in this sense, such as the entanglement entropy and the geometric measure.

These results shed light on the questions addressed in section \ref{sect_background}. For example, an immediate consequence of observation 2 is that a result analogous to theorem 1 will not exist for the entanglement entropy, unless classical and quantum computation have equal power. More generally, these results add to the understanding of which types of entanglement should be generated in large amounts if a quantum algorithm is to yield an exponential speed-up. Perhaps surprisingly and contrary to common intuition, we conclude that, relative to various measures,  it is \emph{not} necessary to generate highly entangled states to achieve universal computational power. 

Whereas for every individual state in the computation the entanglement entropy $E$ (say) will be $\epsilon$-small, the number of repetitions $N$ increases (polynomially) with $1/\epsilon$. Presumably the ``integrated entanglement'' $NE$  will be large. This suggests that this  quantity might be a more appropriate measure of quantum computing power, rather than the entanglement present in any single state during the computation.

The only measures which are not covered by observations 2 and 3 are those that are not sufficiently continuous. Examples are the Schmidt rank and $\alpha$-entropies with $\alpha<1$. Can we hope that the study of this subclass of measures will provide us with comprehensive insights into the (believed) increased power of quantum over classical computation? This is probably unlikely to happen. One immediate dissatisfying feature of such quantities is precisely their discontinuity: one can give examples of quantum circuits where the Schmidt ranks w.r.t certain bipartitions may take on large values (say, scaling linearly with $n$), but where the state of the register remains exponentially close to the trivial state $|0\rangle^n$ throughout the computation. Such computations can be simulated classically in a trivial manner, by simply outputting 0 independent of the details of the circuit. Based on the discussion in section \ref{sect_discontinuous}, similar examples can be given for the $\alpha$-entanglement entropies with $\alpha<1$. Thus, in a trivial sense, such discontinuous measures cannot provide \emph{sufficient} conditions for quantum speed-ups. Therefore such measures will probably not allow us to capture the relationship between quantum and classical computation in a definitive way.

We conclude by raising a question which is naturally stimulated by the present discussion: is it fruitful to adhere to the intuition that pure-state entanglement, as a generic concept with its multitude of manifestations, is a fundamental resource for quantum speed-ups?

\

{\it Acknowledgements.} I thank H. J. Briegel, I. Cirac, M. Christandl, D. Browne, W. D\"ur,  G. Giedke, R. Jozsa,  B. Kraus, M. Piani, R. Renner, G. Vidal and T.-S. Wei for discussions.

\appendix

\section{Basic definitions and notations}\label{sect_app_basic}

The base-2 logarithm of $x$ will be denoted by log $x$. If $A$ is  a matrix with singular values $\{\lambda_i\}$, the trace norm of $A$ is given by \be  \|A\|_1:= \mbox{ Tr} \sqrt{A^{\dagger}A} = \sum |\lambda_i|.\ee The trace distance between two density operators $\rho$ and $\sigma$ is given by \be T(\rho, \sigma):= \frac{1}{2} \|\rho-\sigma\|_1.\ee Note the prefactor $1/2$, which ensures that $T(\rho, \sigma)\leq 1$. The trace distance between two pure states $|\psi\rangle\langle\psi|$ and $|\varphi\rangle\langle\varphi|$ is simply denoted by $T(|\psi\rangle, |\varphi\rangle)$. Pure-state trace distance is related to fidelity as follows:

\be \label{trace_fidelity}T(|\psi\rangle, |\varphi\rangle) = \sqrt{1 - |\langle\psi|\varphi\rangle|^2}.\ee
The operator norm of  $A$ is given by the maximal singular value $\|A\|_{\infty}:= \max |\lambda_i|$. Note that \be\label{trace_operator_norm} |\mbox{Tr}(AB)|\leq \|A\|_{\infty} \|B\|_1\ee for every $A$ and $B$.

\section{Entanglement measures}\label{sect_app_entanglement}

We define the $n$-qubit entanglement measures considered in this paper. We restrict to pure states but most measures have generalizations to mixed states as well. Note that the maximal value of all measures is either $O(n)$ (entanglement entropy, Renyi entanglement entropy, geometric measure, relative entropy of entanglement, squashed entanglement) or $O(1)$  (for the remaining measures). Therefore a scaling of $\delta= $1$/$poly$(n)$, as seen in observations 2 and 3, can for each of these measures be be considered to be ``small''.

There are various distinct notions of what constitutes a valid measure of entanglement (e.g. strong versus weak monotonicity under local operations and classical operations, etc.). In this work such issues will not be important. In fact the only relevant features for our purposes are continuity issues and the fact that entanglement measures vanish on the fully separable state $|0\rangle^n$ (cf. first paragraph of section \ref{sect_continuous}). In the following $|\psi\rangle$ will denote an arbitrary $n$-qubit state.

\

\noindent {\bf Entanglement entropy.} See section \ref{sect_entanglement_entropy}.

\

\noindent {\bf Schmidt rank.} Let $(A, B)$ be a bipartition of the $n$ qubits. Let $\rho^A$ be the reduced density operator of $|\psi\rangle$ for the qubits in $A$. The Schmidt rank $R^{A, B}(|\psi\rangle)$ is given by the rank of the matrix $\rho^A$. In theorem 1 we consider the logarithm of the Schmidt rank: $\chi^{A, B}= \log  R^{A, B}$ .

\

\noindent {\bf Renyi entropies.} See section \ref{sect_discontinuous}.

\

\noindent {\bf Geometric measure \cite{Sh95}.} This measure is defined by \be E_g(|\psi\rangle) = -\log \sup |\langle \psi|\alpha\rangle|, \ee where the optimization is over all complete product states $|\alpha\rangle=|\alpha_1\rangle\otimes\dots\otimes|\alpha_n\rangle$.

\

\noindent {\bf Relative entropy of entanglement \cite{Ve97}.} If $\rho$ and $\sigma$ are $n$-qubit density operators, their quantum relative entropy is defined by \be S(\rho|| \sigma):= \mbox{ Tr}( \rho\log\rho - \rho\log\sigma).\ee The relative entropy of entanglement of an $n$-qubit state $|\psi\rangle$ is given by \be E_{\mbox{\scriptsize{re}}}(|\psi\rangle):= \inf S(|\psi\rangle\langle\psi|\  || \sigma)\ee where the minimization is over all fully separable mixed states $\sigma$.

\

\noindent {\bf Squashed entanglement \cite{Ch04}.} Several different notions of multipartite squashed entanglement exist; however they coincide for pure states. Let $(A_1, \dots, A_m)$ be an arbitrary bipartition of the $n$ qubits into $m$ sets. Let $\rho^{A_i}$ be the reduced density operator of $|\psi\rangle$ for the subset $A_i$. Then the squashed entanglement is given by the sum of the von Neumann entropies \be E_{\mbox{\scriptsize{sq}}}(|\psi\rangle) = \sum_{i=1}^{m} S(\rho^{A_i}).\ee

\

\noindent {\bf Localizable entanglement \cite{Ve04}.} This measure is defined as the maximal amount of entanglement that can be created, on average, between qubits $i$ and $j$ by performing
local measurements on the other qubits. Consider an arbitrary  protocol ${\cal L}$ of single-qubit projective measurements on the qubits outside $i$ and $j$ which transforms $|\psi\rangle$ into $K$ possible output states, each with some probability $p_{\mu}$. Let $|\psi^{\mu}_{ij}\rangle$ denote the state of qubits $i$ and $j$ in branch $\mu$ of the protocol, where $\mu=1\cdots K$. The average entanglement generated in the protocol ${\cal L}$ is \be E^{\cal L}_{ij}(|\psi\rangle) = \sum p_{\mu} E(|\psi^{\mu}_{ij}\rangle),\ee where $E$ denotes the bipartite entanglement entropy. The localizable entanglement is then the supremum value over all such protocols: \be LE_{ij}(|\psi\rangle) = \sup_{\cal L} E^{\cal L}_{ij}(|\psi\rangle).\ee

\

\noindent {\bf Multipartite concurrence.} The multipartite concurrence generalizes the standard 2-qubit concurrence. Let ${\cal H}$ denote a 2-qubit Hilbert space and let $P$ denote the projector onto the symmetric subspace of ${\cal H}$. The multipartite concurrence of the $n$-qubit state $|\psi\rangle$ is  \footnote{The multipartite concurrence is usually defined in a different way but Eq. (\ref{concurrence}) was shown to be equivalent \cite{Ao06}.} \be\label{concurrence} C(|\psi\rangle):= 2 \sqrt{1- \langle\psi|^{\otimes 2} P_1\otimes \dots \otimes P_n|\psi\rangle^{\otimes 2}}.\ee Here $P_i$ denotes the operator $P$ acting on the $i$-th qubits of both copies of $|\psi\rangle$.

\

\noindent {\bf n-Tangle}. The $n$-tangle is another multipartite generalization of the concurrence, defined for $n$ qubits with even $n$ by \be N(|\psi\rangle) = |\langle\psi|Y^{\otimes n}|\psi^*\rangle|^2\ee where $|\psi^*\rangle$ denotes the state obtained by complex conjugating the coefficients of $|\psi\rangle$ in the computational basis, and where $Y$ denotes the standard $\sigma_y$ Pauli operator.

\section{Proof of observation 3}

We show that for each of the measures $E$ appearing in (a)-(g) a suitable polynomially small $\epsilon$ can be chosen such that $E(|\psi\rangle)=O(\delta)$ for every $|\psi\rangle\in {\cal S}_{\epsilon}$. In combination with observation 1, this will prove observation 3. In the following we let $|\psi\rangle\in {\cal S}_{\epsilon}$.

\

\noindent (a) The claim is proved by using observation 2 and the property that $E_{\alpha}^{A, B}(|\psi\rangle)\leq E^{A, B}(|\psi\rangle)$ for every $\alpha\geq 1$ and for every bipartition $(A, B)$.

\

\noindent (b) Using (\ref{trace_fidelity}) one has $|\langle \psi|0\rangle^n|= \sqrt{ 1-\epsilon^2}$. This implies that  \be E_g(|\psi\rangle) \leq  -\log \sqrt{1-\epsilon^2}= O(\epsilon^2). \ee

\

\noindent (c) Some care is required since the quantum relative entropy may be infinitely large. In fact $S(\rho|| \sigma) = \infty$ whenever $\rho$ and $\sigma $ are distinct pure states. Thus in particular generally $S(|\psi\rangle || |0\rangle^n) = \infty$ so that the ``obvious'' approach to prove that $E_{re}(|\psi\rangle)\leq \delta$ does not work. Nevertheless this problem can easily be resolved by replacing $|0\rangle^n$ with a full-rank state $\sigma$ which is close to $|0\rangle^n$, as shown next.

We will use the following continuity bound from Ref. \cite{Au05}. Let $\rho$ and $\sigma$ be two $n$-qubit density operators, let $T$ denote their trace distance and let $\lambda$ be the minimum eigenvalue of $\sigma$. Suppose furthermore that $-2T \log 2T\leq 1/e$.  Then \be\label{cont_relent} S(\rho|| \sigma)  \leq 2 n T  -2T \log 2T - T \log \lambda.\ee Now consider $\rho:= |\psi\rangle\langle\psi|$ and \be \sigma:= (1-\epsilon) |0\rangle\langle0|^n + \epsilon M\ee where $M := I/2^n$ is the fully mixed state on $n$ qubits. Remark that $\sigma$ is a separable state. It is easily verified that the smallest eigenvalue of $\sigma$ is $\lambda:= \epsilon/2^n$. Furthermore, using that $T(|\psi\rangle, |0\rangle^n)\leq \epsilon$ and $T(|0\rangle^n, \sigma)\leq \epsilon$ and the triangle inequality, it follows that \be T( \rho, \sigma) \leq T( \rho, |0\rangle^n)  + T(|0\rangle^n, \sigma)\leq 2 \epsilon.\ee Using (\ref{cont_relent}) it follows that \be  S(\rho|| \sigma) =O(n\epsilon). \ee By choosing $\epsilon$ sufficiently polynomially small, one can ensure that $S(\rho|| \sigma)= O(\delta)$ for every polynomially small $\delta$. But then $E_{\mbox{\scriptsize{re}}}(|\psi\rangle) =O(\delta)$ as well.

\

\noindent (d) The proof follows from observation 2.

\

\noindent (e) The proof follows from observation 2 and the property that $ LE_{ij}(|\psi\rangle)$ is not greater than the entanglement entropy of $|\psi\rangle$ w.r.t. the bipartition $(\mbox{qubit } i, \mbox{ rest})$.

\

\noindent (f), (g):  see appendix \ref{sect_app_polynomial}.

\section{Entanglement measures based on polynomial functions}\label{sect_app_polynomial}

Consider a function \be\label{f_polynomial} f(|\psi\rangle) = \langle \psi|^{\otimes k} A |\psi\rangle^{\otimes k}\ee where $A$ is an $nk$-qubit (possibly non-Hermitian) operator. This function represents a homogeneous polynomial of degree $k$ in the matrix entries of the density operator $\rho=|\psi\rangle\langle\psi|$. In fact every such polynomial can be written in the form (\ref{f_polynomial}).

Let $|\psi\rangle$ and $|\varphi\rangle$ be $n$-qubit states and denote their trace distance by $T$.  We claim that \be\label{claim_f} |f(|\psi\rangle) - f(|\varphi\rangle)|\leq  4 k T \|A\|_{\infty}. \ee To prove this, let $T_k$ denote the trace distance between $|\psi\rangle^{\otimes k}$ and $|\varphi\rangle^{\otimes k}$. Using (\ref{trace_operator_norm}) yields \be |f(|\psi\rangle) - f(|\varphi\rangle)| &=& \left| \mbox{Tr} \left\{ A \left[|\psi\rangle\langle\psi|^{\otimes k} - |\varphi\rangle\langle\varphi|^{\otimes k} \right] \right\} \right|\nonumber\\ &\leq& 2 \|A\|_{\infty}T_k.\ee Owing to (\ref{trace_fidelity}) we have \be T_k^2 = 1 - |\langle\psi|\varphi\rangle|^{2k}= 1 - (1- T^2)^{2k}.\ee Using that $(1-x)^{n}\geq 1-n^2 x$ for every  integer $n\geq 2$ and for every $x\in[0, 1]$ (which is easily proved using the binomial expansion), we find $T_k \leq 2k T$. This yields (\ref{claim_f}).

Now consider a family of functions $f_n$ of the form (\ref{f_polynomial}) where $n=1, 2, \dots$ and where $f_n$ acts on $n$-qubit states with associated $k_n$ and $A_n$. Assume that  both $k_n$ and $\|A_n\|_{\infty}$ scale at most polynomially with $n$. Then (\ref{claim_f}) implies the following:

\

\noindent {\it Property 1.} For every polynomially small $\delta_n$ there exists a polynomially small $\epsilon_n$ such that \be |f_n(|\psi\rangle) - f_n(|\varphi\rangle)|=O(\delta_n)\ee holds for all $n$-qubit states $|\psi\rangle$ and $|\varphi\rangle$ satisfying $T(|\psi\rangle, |\varphi\rangle)\leq \epsilon_n$.

\

The above basic continuity property of polynomial functions can  be used to show that a number of entanglement measures are continuous in the sense required for observation 3, since several measures are given in terms of a suitable $f$ or as simple functions thereof (e.g. square roots or linear combinations of a few such expressions). For example property 1 can readily be used to prove observation 3(f) by remarking that $C(|\psi\rangle) = 2\sqrt{1-f(|\psi\rangle)}$ for some suitable $f$ of the form (\ref{f_polynomial}) with $k=2$ and $\|A\|_{\infty} = 1$ since $P$ is unitary.

Property 1 can also be applied to witness-based measures (\ref{E_witness}).  Consider an $n$-qubit measure \footnote{More, precisely, we are considering a family of measures $E_{{\cal C}_1}, E_{{\cal C}_2}, \dots$ where $E_{{\cal C}_n}$ is defined on $n$-qubit states.} $E_{{\cal C}}$ and assume that the norm of each witness is polynomially bounded i.e. there exists a polynomial $p(n)$ such that \be\label{W_norm1} \|W\|_{\infty}= O(p(n))\mbox{ for every } W\in {\cal C}.\ee Then $E_{{\cal C}}$ will be continuous in the sense considered in observation 3. To see this, consider $|\psi\rangle\in {\cal S}_{\epsilon}$ as before.   
Then
\be\label{W_norm2}  - \langle\psi|W|\psi\rangle &\leq& -\langle\psi|W|\psi\rangle + \langle 0|^n W|0\rangle^n\nonumber\\
 &\leq& |\langle\psi|W|\psi\rangle - \langle 0|^nW|0\rangle^n|\nonumber\\ &\leq& 2\epsilon \|W\|_{\infty}\ee for every $W\in {\cal C}$. In the first inequality we used that $\langle 0|^n W|0\rangle^n\geq 0$ since  $W$ is an entanglement witness and $|0\rangle^n$ is a product state; in the second inequality we used that $x\leq |x|$ for all real $x$; in the third inequality we used (\ref{trace_operator_norm}). Equation (\ref{W_norm1}) with (\ref{W_norm2}) implies that \be E_{{\cal C}}(|\psi\rangle)=O(\epsilon p(n)).\ee Therefore, for every polynomially small $\delta$ there exists a polynomially small $\epsilon$ ensuring that $E_{{\cal C}}=O(\delta)$.

An analogue of property 1 can also be proved for functions of the form
\be\label{g_polynomial} g(|\psi\rangle) = \langle \psi|^{\otimes k} A |\psi^*\rangle^{\otimes k}\ee where $|\psi^*\rangle$ denotes the state obtained by complex conjugating the coefficients of $|\psi\rangle$ in the computational basis. We have \be &&\| |\psi\rangle\langle\psi^*|^{\otimes k}  - |\varphi\rangle\langle\varphi^*|^{\otimes k} \|_1  \nonumber\\ &\leq& \| |\psi\rangle\langle\psi^*|^{\otimes k}  - |\varphi\rangle\langle\psi^*|^{\otimes k}\|_1 + \|  |\varphi\rangle\langle\psi^*|^{\otimes k} - |\varphi\rangle\langle\varphi^*|^{\otimes k} \|_1\nonumber\\ &\leq& \| |\psi\rangle^{\otimes k}  - |\varphi\rangle^{\otimes k}\|_1 + \|  |\psi^*\rangle^{\otimes k} - |\varphi^*\rangle^{\otimes k} \|_1 \nonumber \\ &=& 2\| |\psi\rangle^{\otimes k}  - |\varphi\rangle^{\otimes k}\|_1.\ee In the first inequality we used the triangle identity, in the second inequality we used that $\|AB\|_1\leq \|A\|_1\|B\|_1$ and in the following identity we used that $\|A^*\|= \|A\|$. With this, using an argument analogous to above one finds that
\be\label{claim_g} |g(|\psi\rangle) - g(|\varphi\rangle)|\leq  8 k T \|A\|_{\infty}.\ee This readily yields an analogue of property 1 for the functions $g$. The latter can e.g. be used to prove observation 3 for the $n$-tangle.



\begin{thebibliography}{99}







\bibitem{Jo97}
R. Jozsa,  quant-ph/9707034 (1997).

\bibitem{Go98}
D. Gottesman, arXiv:quant-ph/9807006 (1998).


\bibitem{Li01}
N. Linden and S. Popescu, Phys. Rev. Lett. 87, 047901 (2001).



\bibitem{Jo03}
R. Jozsa and N. Linden, Proc. R. Soc. Lond. A vol. 459(2036), pp. 2011-2032 (2003).

\bibitem{Vi03}
G. Vidal, Phys. Rev. Lett. 91, 147902 (2003).



\bibitem{Or04}
R. Orus and J. I. Latorre, Phys. Rev. A 69, 052308 (2004).


\bibitem{Bi04}
E. Bihama, G. Brassard, D. Kenigsberg and T. Mora, Theoretical Computer Science Vol 320(1) pp. 15–33 (2004).

\bibitem{Da05}
A. Datta, S. T. Flammia, and C. M. Caves, Phys. Rev. A 72, 042316 (2005).

\bibitem{Va06}
M. Van den Nest, A. Miyake, W. D\"ur, and H. J. Briegel, Phys. Rev. Lett 97 150504 (2006).

\bibitem{Jo06}
R. Jozsa, quant-ph/0603163 (2006).

\bibitem{Va07}
M. Van den Nest, W. D\"ur, G. Vidal and H.J. Briegel, Phys. Rev. A 75, 012337 (2007).

\bibitem{Ma08}
I.L. Markov and Y. Shi, SIAM J. Comp. 38(3), 963 (2008).






\bibitem{Yo08}
N. Yoran, arXiv:0802.1156 (2008).




\bibitem{Gr09}
D. Gross, S. T. Flammia and J. Eisert, Phys. Rev. Lett. 102, 190501 (2009).













\bibitem{Sh95}
A. Shimony, Ann. N.Y. Acad. Sci. 755, 675 (1995); H. Barnum and N. Linden, J. Phys. A 34, 6787 (2001); T.-C. Wei and P. M. Goldbart, Phys. Rev. A 68, 042307 (2003).

\bibitem{Ve04}
F. Verstraete, M.-A. Martin-Delgado and J.I. Cirac, Phys. Rev. Lett. 92, 087201 (2004).

\bibitem{Ve97}
V. Vedral, M.B. Plenio, M.A. Rippin, and P.L. Knight, Phys. Rev. Lett. 78, 2275 (1997).


\bibitem{Ch04}
M. Christandl and A. Winter, J. Math. Phys 45, 829 (2004); Yang et al., IEEE Trans. Inf. Theory 55, 3375 (2009).


\bibitem{Ca04}
A. R. R. Carvalho, F. Mintert, A. Buchleitner, Phys. Rev. Lett
93, 230501 (2004).

\bibitem{Pi08}
M. Piani, C. Mora and H. J. Briegel, New J. Phys. 10, 083027 (2008).

\bibitem{Wo01}
A. Wong and N. Christensen, Phys. Rev. A 63, 044301 (2001).

\bibitem{Fa73}
M. Fannes, Commun. Math. Phys. 31 291 (1973).




\bibitem{Va02}
L. G. Valiant, SIAM J. Comput. 31, No. 4, p. 1229 (2002); D. DiVincenzo and B. Terhal, Phys. Rev. A 65, 032325/1-10 (2002); R. Jozsa and A. Miyake, Proc. R. Soc. A 464, 3089-3106 (2008); M. Van den Nest, Quant. Inf. Comp. 11, 9-10 pp. 784-812 (2011); M. Van den Nest, arXiv:1201.4867 (2012).





\bibitem{Ao06}
L. Aolita and F. Mintert, Phys. Rev. Lett. 97, 050501 (2006).





\bibitem{Bu11}
F. Buscemi and N. Datta, Phys. Rev. Lett. 106, 130503 (2011).

\bibitem{Os05}
A. Osterloh and J. Siewert, Phys. Rev. A 72 012337 (2005).

\bibitem{Br05}
F. G. S. L. Brandao,  Phys. Rev. A A 72, 022310 (2005).


\bibitem{Ve06}
F. Verstraete and J. I. Cirac, Phys. Rev. B 73, 094423 (2006).


\bibitem{Ei01}
J. Eisert and H. J. Briegel, Phys. Rev. A 64, 022306 (2001)



\bibitem{Au05}
K. M. R. Audenaert and J. Eisert, J. Math. Phys. 46, 102104 (2005)

\end{thebibliography}
\end{document}